\documentclass[preprint]{aastex}

\usepackage{graphicx}

\slugcomment{Accepted for publication in PASP:  May 2007}

\shorttitle{Water Vapor at Las Campanas Observatory}
\shortauthors{Thomas-Osip et al.}

\begin{document}

\title{Calibration of the Relationship between Precipitable Water Vapor and 
225~GHz Atmospheric Opacity via Optical Echelle Spectroscopy at Las Campanas 
Observatory}

\author{Joanna Thomas-Osip,\altaffilmark{1} Andrew McWilliam, M.~M. Phillips\altaffilmark{1}, N. Morrell\altaffilmark{1} and I. Thompson}
\affil{The Observatories of the Carnegie Institute of Washington\\
813 Santa Barbara St, Pasadena, CA 91101}
\email{ jet@lco.cl, andy@ociw.edu, mmp@lco.cl, nmorrell@lco.cl, ian@ociw.edu}

\author{T. Folkers}
\affil{Arizona Radio Observatory, University of Arizona,Tucson AZ 85721, USA}
\email{tfolkers@hamms.as.arizona.edu}

\author{T. Folkers}
\affil{Physics Department and Astronomy Department\\
University of Michigan, Ann Arbor, MI 48109}
\email{fca@umich.edu}

\and

\author{M. Lopez-Morales}
\affil{Carnegie Institution of Washington, Department of Terrestrial Magnetism\\
5241 Broad Branch Road, NW, Washington, DC 20015}
\email{mercedes@dtm.ciw.edu}

\altaffiltext{1}{Las Campanas Observatory, Colina El Pino, Casilla 601, La Serena, Chile}

\begin{abstract}

We report precipitable water vapor (PWV) measurements made at Las Campanas
Observatory using optical spectra of H$_2$O lines obtained with the Magellan 
echelle spectrograph, and calculated using a robust technique that is accurate to 
5-10\%.
Calibration of the relationship between our PWV measurements and opacity values at 
225~GHz was made possible by simultaneous tipping radiometer observations.  Based on this
calibration, we present Las Campanas Observatory winter-time precipitable water vapor 
statistics, measured using the tipping radiometer, during a two month campaign.  The 
median value of 2.8$\pm$0.3 mm is consistent with that measured at the nearby La Silla
Observatory during the VLT site survey.
We conclude that, in the Southern hemisphere winter months, we can expect good conditions 
for infrared observing ($\lesssim$1.5 mm) approximately 10\% of the time at Las Campanas 
Observatory.
\end{abstract}


\section{Introduction}

As part of the Giant Magellan Telescope site testing program at Las Campanas Observatory, 
we are interested in characterizing precipitable water vapor due to its impact on mid-IR 
observations through decreased transparency, increased thermal-IR background, and the 
introduction of extraneous spectral features.  In this study, we report on our efforts to 
absolutely calibrate the relationship between 225~GHz opacity and precipitable water vapor (PWV).

The opacity at 225~GHz has been shown to be directly related to the column of precipitable
water vapor in the atmosphere \citep[c.f.][]{pla78,cb95,dav97}.  The relationship, however,
is also a function of pressure (and therefore altitude) and temperature due to the fact that the 
225~GHz opacity derives from the wing of a strong water vapor absorption line at 183~GHz 
arising from the 3-2 rotational transition \citep{wat76} where the wings are mostly due to
pressure broadening.  Efforts to calibrate this relationship have taken several different 
forms including deducing the column of PWV from a model of the atmospheric transmission at
mm wavelengths and measuring the PWV using radiosonde type devices.  Comparison between 
methods can be complicated by the effects of other atmospheric constituents (e.g. $O_3$, 
$O_2$, and $N_2$) on the dry air opacity.

\citet{swi90} combined high resolution echelle spectroscopy of a water vapor 
line at 6943.79~\AA\ with a technique devised by \cite{b75} (henceforth B75) 
to calibrate mid-infrared sky radiance measurements obtained at the La Silla 
Observatory during the VLT site survey.  Through the use of modern partition 
functions and the addition of 13 new optical lines for PWV measurement, we have
improved substantially on the B75 method when used in conjunction with high 
resolution echelle spectroscopy.  Section~\ref{sec-opt} describes the method 
in detail and our results using it.  Our campaign to measure the 225~GHz 
opacity using a tipping radiometer, the resulting calibration and PWV 
statistics for Las Campanas Observatory are presented in \S~\ref{sec-tip}.

\section{Optical Echelle Spectroscopy}\label{sec-opt}

The Magellan echelle spectrograph (MIKE) was used to acquire spectra of
rapidly rotating A and B stars in the visual magnitude range 4.0 to 6.0.
Due to the high continuous opacity these hot stars show very few absorption lines,
particularly in the red optical region.  Furthermore, their
high rotational velocities ensure that any extant stellar lines are rotationally
broadened, to widths much larger than lines arising from absorption through the
earth's atmosphere.

The Las Campanas Observatory site monitoring program includes a list of 16 telluric 
standard stars\footnote{http://www.lco.cl/lco/operations-inf/gmt-site-testing-1/stars-for-measuring-pwv-with-mike/stars-for-measuring-pwv}
that pass close to the zenith, such that at any time one, or more, stars are within
2 hours of the meridian.  Observers were instructed not to submit echelle spectra taken
through cirrus clouds as a precaution against spurious PWV measurements arising from
clumpy clouds.

Typically, spectra were acquired when the hot stars were close to the zenith.
The spectral resolving power was normally $\sim$40,000 with S/N in excess of 100:1.
The echelle spectral format offers the advantage that light of a given wavelength
occurs in two consecutive orders.  This permits independent measurement of each line
in two orders, providing that the lines fall on the echelle CCD array and there
is sufficient signal.

Reductions of the raw spectra were performed using the MIKE Pipeline reduction
software, written by Dan Kelson, using algorithms outlined in Kelson et al. (2000),
Kelson (2003) and Kelson (2006).

While the site survey program, including the measurement of PWV, is on-going,
this paper is based on measurements of 15 spectra, taken on 15 nights
from 21 July to 21 November 2005.  The Tipper was not operational on every
night that echelle spectra were taken; thus, our Tipper calibration is based
on 11 of the 15 spectra.

\begin{figure}
\plotone{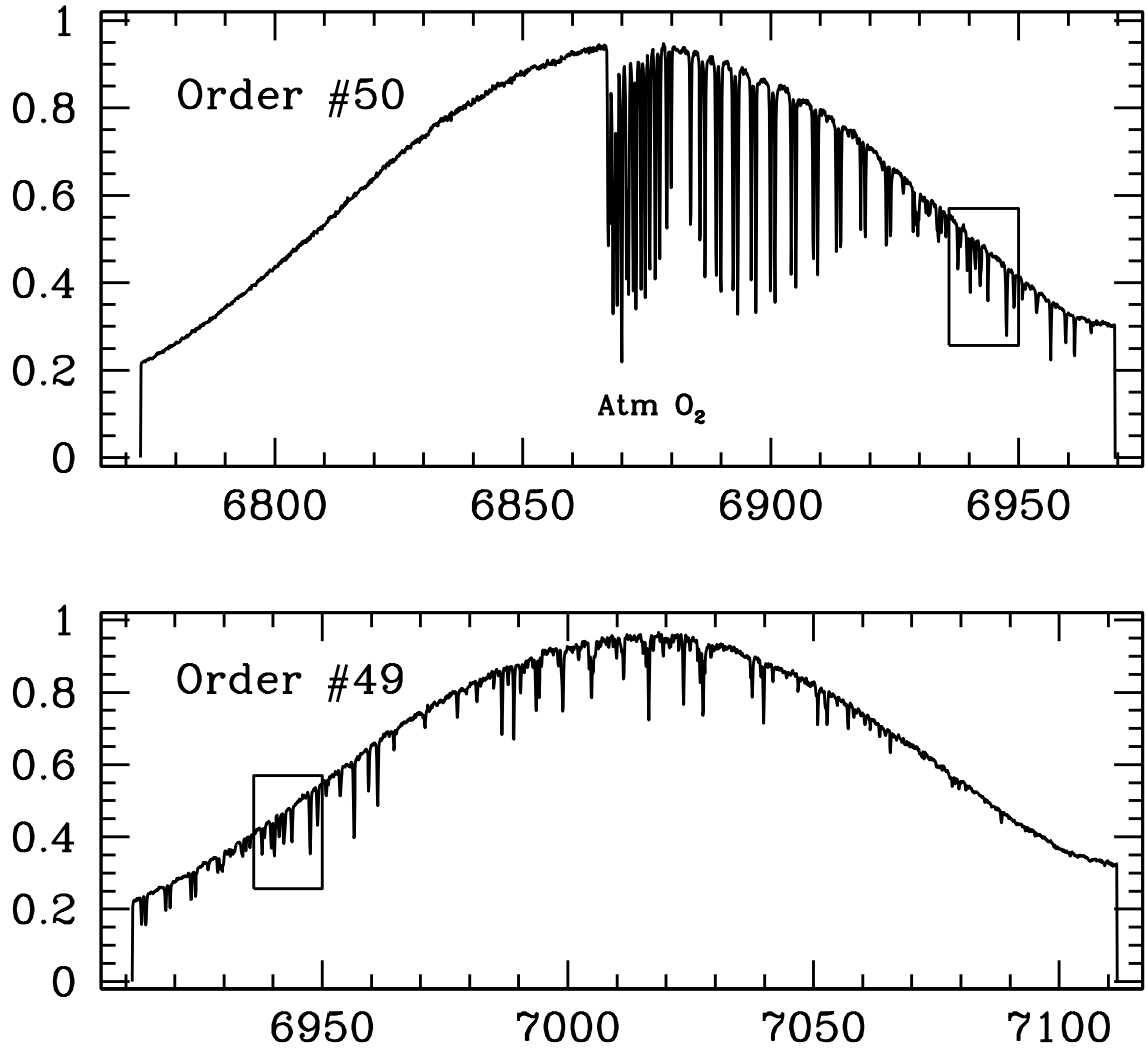}
\caption{Two consecutive orders containing the H{$_2$}O line at 6943.78~\AA ,
for a night with 7mm PWV.  Boxes indicate the regions shown in detail in the next figure.}
\label{fig-bluered}
\end{figure}

\begin{figure}
\plotone{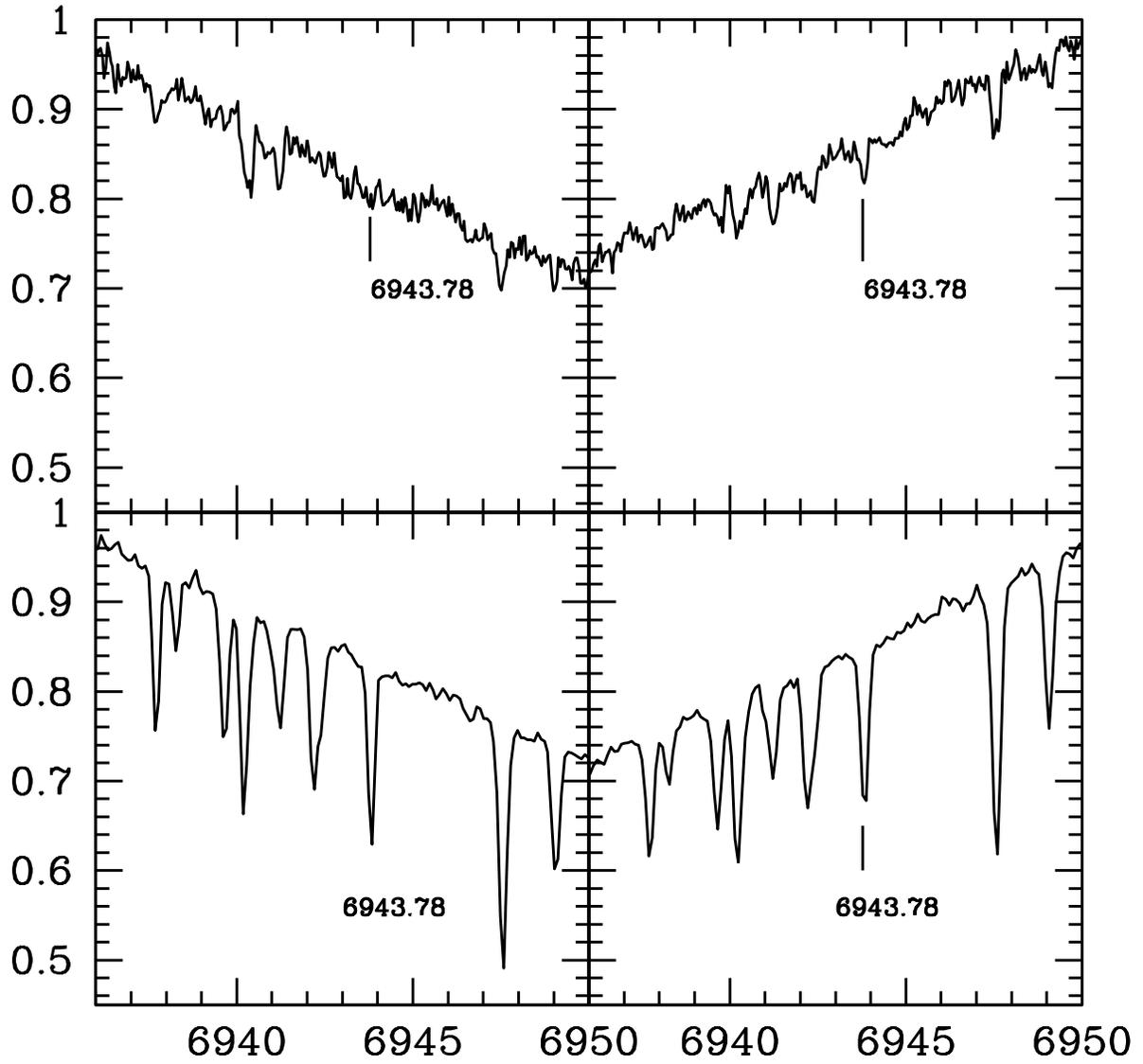}
\caption{A comparison of the 6943.78~\AA\ H{$_2$}O line in the two consecutive orders for
a dry night, with PWV {$\sim$}1.0mm, and a wet night with PWV {$\sim$}7.0mm.}
\label{fig-drywet}
\end{figure}

Figures~\ref{fig-bluered} and \ref{fig-drywet} show the spectrum near the H$_2$O line
used by B75 at 6943.79~\AA, in consecutive orders on our driest and wettest nights.  
The equivalent widths of the line in these spectra ranges from 5 to 52~m\AA .

\subsection{Analysis of the Spectra}\label{sec-ana}

Following  B75, we analyze our spectra using a robust and computationally simple method, 
but we employ improved partition functions that result in a slightly different choice of 
line excitation potentials than B75.  We have also added five new weak optical lines that 
may be used for reliable PWV estimates.

B75 pointed out that for a given temperature there is a certain, unique, energy level
which corresponds to a temperature-insensitive absorption coefficient.
This follows from the Boltzmann equation (Equation~1) 
describing the population of the energy levels of atoms and molecules:
at any given temperature, T, the population of levels with a unique excitation potential
matches the sensitivity of the partition function to temperature; in this case the
fractional population of the level (n$_i$/N$_{\rm tot}$) is approximately constant with T,

\begin{equation}
\frac{n_i}{N_{Tot}} = \frac{g_i e^{-EP_i/kT}}{Q(T)}
\end{equation}

\noindent where the partition function, Q(T), is given by

\begin{displaymath}
Q(T) = \sum g_j e^{-EP_j/kT} 
\end{displaymath}

\begin{figure}
\plotone{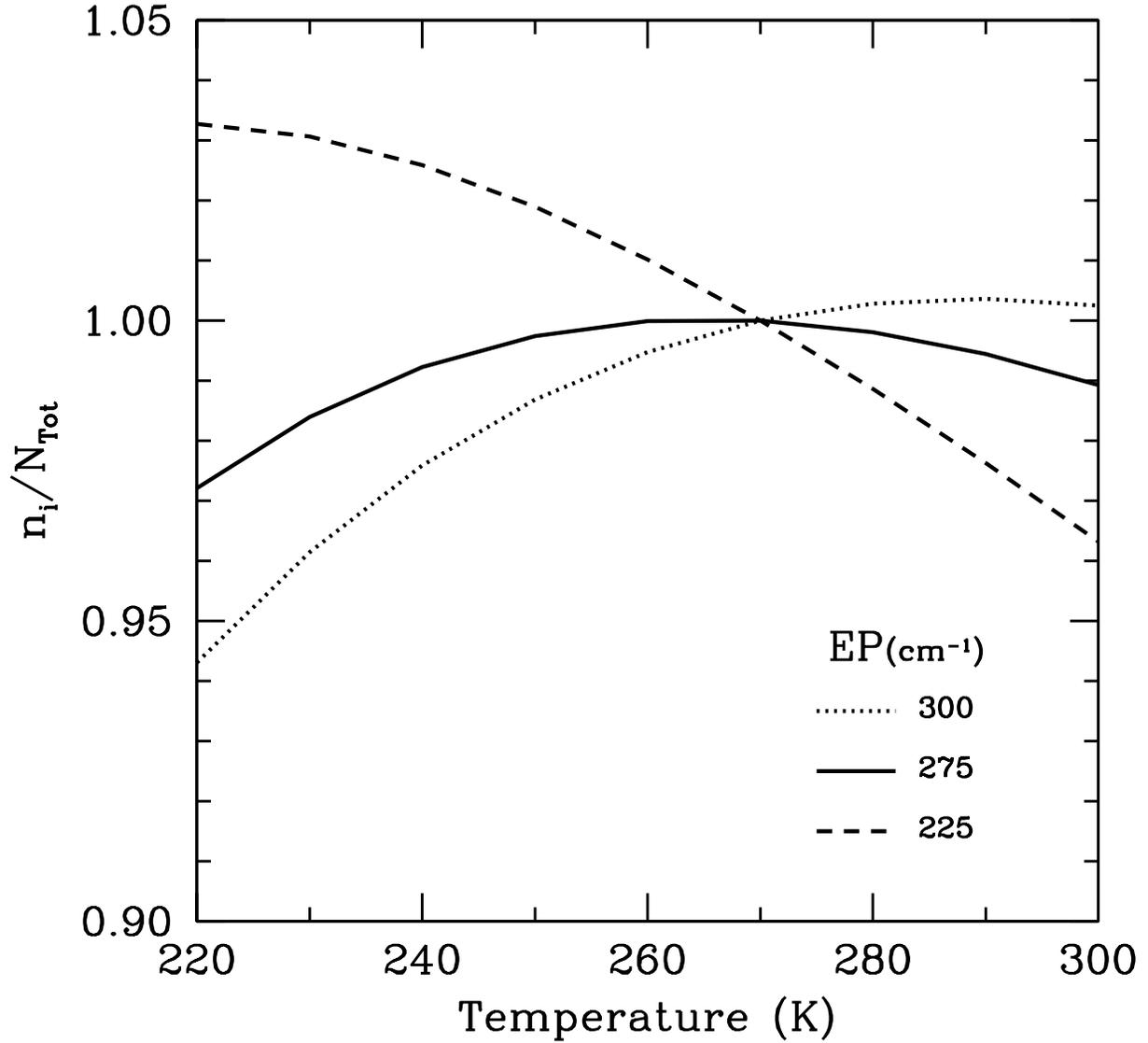}
\caption{A plot showing the fractional population of levels (n$_i$/N$_{\rm tot}$)
in H$_2$O as a function of temperature, in the range 220--300~K, for level energies
of 225, 275 and 300 cm$^{-1}$.  Over the range of temperatures seen in the earth's
atmosphere the fractional population of levels in this energy level range changes
very little, and allows for robust measurement of PWV, without a detailed model atmosphere.}
\label{fig-bfac}
\end{figure}

In Figure~\ref{fig-bfac} we show a plot of (n$_i$/N$_{\rm tot}$) versus temperature for 
three different
energy levels, over a range of temperatures expected in the earth's atmosphere; the
plots were calculated using Equation~1, and are normalized to a temperature of T=270~K.
Figure~\ref{fig-bfac} is similar to Figure~1 in B75, but here we have used the 
partition function taken from the
Kurucz\footnote{http://kurucz.harvard.edu/} website that we believe is more appropriate 
than the simple approximation for the H$_2$O partition function employed by B75.  
Figure~\ref{fig-bfac} indicates
that for levels with excitation energy in the range 225--300 cm$^{-1}$ the
ratio (n$_i$/N$_{\rm tot}$) varies by less than $\sim$5\% over the temperature range
from 220 to 300~K.  The excitation energy range
least sensitive to atmospheric temperature, around 270--280 cm$^{-1}$, is slightly
higher than the value of 225 cm$^{-1}$ indicated by B75, and probably results from our
use of modern partition functions.  We note that even the $\sim$5\% uncertainty could
be reduced by use of a weighted average of the results from lines 
spanning the critical range of excitation energies.  

\begin{figure}
\plotone{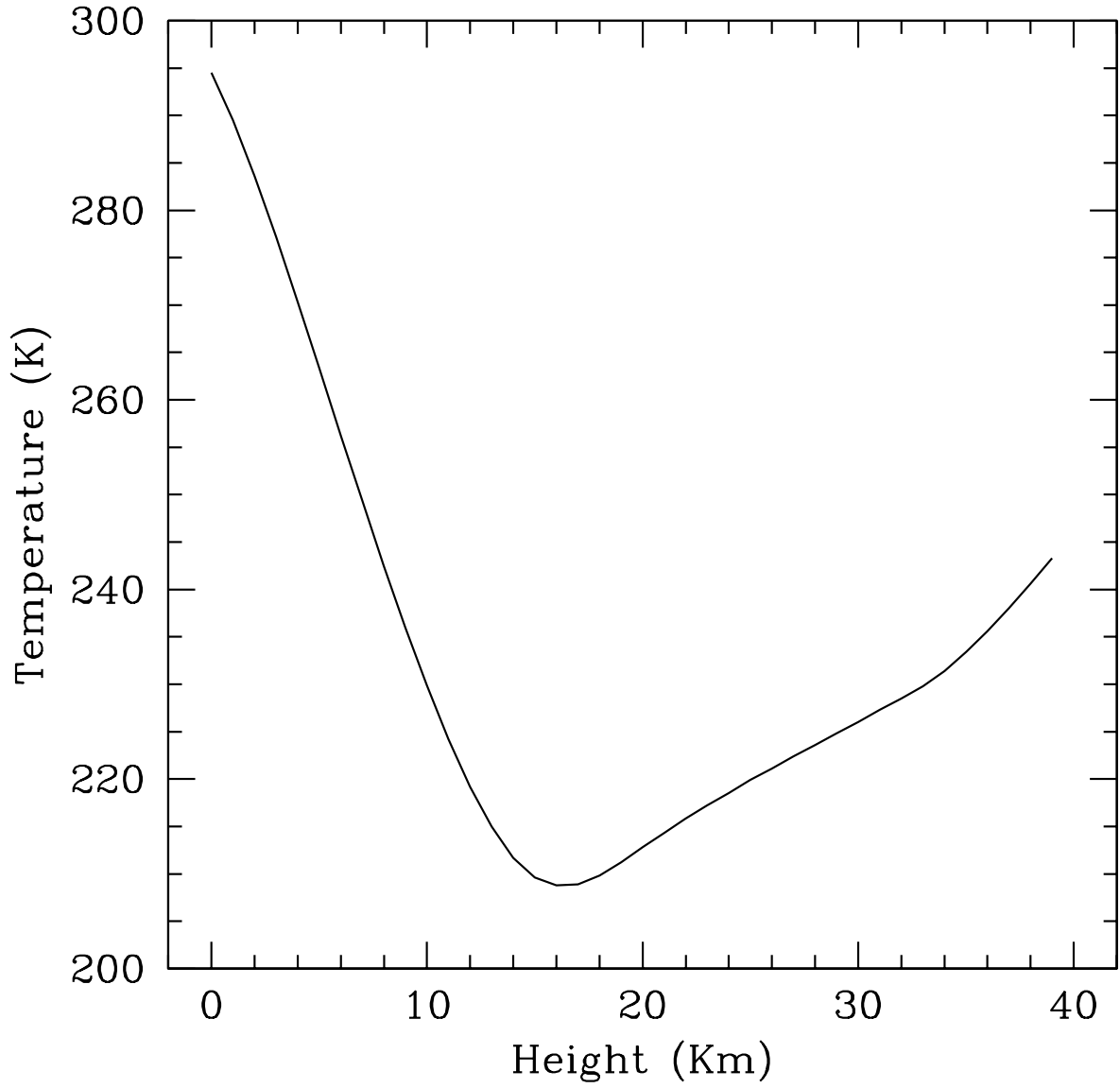}
\caption{MSIS-E-90 model atmosphere}
\label{fig-atmos}
\end{figure}

Figure~\ref{fig-atmos} shows a 
MSIS-E-90\footnote{http://modelweb.gsfc.nasa.gov/models/msis.html} model atmosphere
for Las Campanas Observatory on one of the nights for which we have a PWV
spectrum; it can be seen that the temperature range of the model is covered by the
range in Figure~\ref{fig-bfac}.  The largest temperature deviation from 270~K, in the model, 
is 210~K and occurs at the tropopause
between 10 and 20~km height; it encompasses approximately 20\% of the mass of
the atmosphere.  We note that the water vapor pressure above ice, or water, at 
220~K is approximately 0.6\% of the vapor pressure at 270~K (e.g. Mason 1971).  Thus, if the
H$_2$O is uniformly distributed through the atmosphere we estimate
that approximately 0.1\% of the mass of the atmospheric water vapor 
lies in the 10--20~km region, where the temperature differs most from 270~K.  
In other words the coolest region of the atmosphere contains a negligible fraction
of the total water vapor, and thus even the small systematic deviation of
n$_i$/N$_{\rm tot}$ from this region will be reduced to insignificance compared to the
whole atmosphere.  We conclude that lines with excitation levels near 270--280 cm$^{-1}$ 
can provide very robust estimates of the atmospheric PWV.

Under the assumption of pure absorption radiative transfer, and a constant 
line profile function though the earth's atmosphere, B75 pointed out that the
PWV, in cm, is given by:

\begin{equation}
PWV = L /SX
\end{equation}

\noindent where L is the log flux, 
S is the line strength parameter, and X is the airmass.  B75 referred to
L as the ``log equivalent width'', and defined it as the
the integral of the natural log of the flux removed from the continuum
by the line:

\begin{displaymath}
L = \int^{\infty}_{-\infty} -log_e \left(\frac{I(\nu)}{I_c}\right)d\nu
\end{displaymath}

\noindent The line strength parameter S depends on atomic parameters as
follows:

\begin{equation}
S =  gf \left(\frac{\pi e^2}{mc}\right)  \frac{e^{-EP/kT}}{Q(T)}  \left(\frac{N_A}{A_{H_2O}}\right)
\end{equation}

\noindent Note that the $gf$ value can be calculated from the Einstein 
spontaneous emission coefficient, $A_{ul}$, by the following relation:

\begin{displaymath}
gf = 1.4992 \lambda^2 g_{up} A_{ul}
\end{displaymath}

\noindent where $\lambda$ is in cm and $A_{ul}$ is in Hertz.

In Equation~3 $N_A$ is Avogadro's number, 6.022$\times$10$^{23}$, and 
$A_{H_2O}$ is the atomic mass of H$_2$O, at 18.0.  Note that use of 
Equation~3 will give PWV, in cm, if the log flux integration, L, is performed 
over the frequency spectrum of the line in Hertz units.  For an assumed 
temperature of 270~K, which is our single model atmosphere parameter, 
Equation~3 reduces to:

\begin{equation}
S_{270} = \frac{2.917335\times10^8}{\sigma_{vac}^2}  A_{ul}  g_{up}  e^{-EP/187.6524}
\end{equation}

\noindent In Equation~4 we have applied the additional factor 
$-c\sigma_{vac}^2$ 
(where $\sigma_{vac}$ is the vacuum wavenumber) in the calculation of
the strength factor, S, to account for the line flux measured in Kaysers 
(cm$^{-1}$), consistent with the S factors given in B75.

In practice the line profile depends most strongly on the atmospheric pressure, 
which is a function of height, so a constant line profile condition, mentioned by B75,
does not apply.  However, in the case of unsaturated lines, the line profile does not 
affect the total absorbed flux, and Equation~2 becomes valid, even with large 
variations in the line profile through the atmosphere.  In this study we 
computed PWV values from our measured H$_2$O line log fluxes, with wavelengths 
in cm$^{-1}$ units, using Equations~2 and~4; this is correct for unsaturated 
lines.  For saturated lines the variability of the line profile with height 
signals the break-down of Equation~2.  In this case, to properly measure PWV 
would require a model atmosphere that accounts for the relation between line 
profile and column mass of H$_2$O.  

B75 measured the line strength parameter, S, using the McMath Solar Telescope 
for numerous lines with excitation near 225 cm$^{-1}$, mostly for lines in the 
near-infrared region of the spectrum.  However, B75's list did include one 
line in the optical, at 6943.79~\AA.  As a check of the B75 value of S for 
this line, we computed S from the Kurucz\footnote{http://kurucz.harvard.edu/} 
$gf$ value using Equation~4.  The Kurucz $gf$ value, which was taken from 
theoretical predictions of Partridge \& Schwenke (1997), leads to a value of
S which is $\sim$60\% higher than the B75 measurement, although there is
reason to suspect the veracity of this result since the relative Kurucz 
$gf$ values appear to be inconsistent with the appearance of lines in our 
telluric spectra.

Given this disagreement, we decided to check the result from the 
6943.79~\AA\ line by identifying additional optical H$_2$O lines with 
excitation energies near 270--280 cm$^{-1}$ and reliable, measured, $gf$ 
values.  Our criteria for selecting the additional lines also included the 
requirement that they have a strength that could be measured in our spectra, 
but not so strong that they might be saturated, and that the lines be clear 
from blends, including H$_2$O lines and other lines present in our telluric 
spectra (e.g. from the O$_2$ molecule).   We searched the HITRAN 
\footnote{http://cfa-www.harvard.edu/HITRAN/} database of H$_2$O line 
parameters for suitable lines and visually inspected our telluric spectra to 
check for non H$_2$O blends.

In Table~\ref{tab-params} we show line parameters for the 6943.79~\AA\ line of B75 and 5 
new weak lines identified in this work.  The HITRAN database included optical H$_2$O line 
$gf$ values from a variety of sources.  The $gf$ values for our lines in 
Table~\ref{tab-params} came from Coheur et al. (2002), measured using a high resolution 
spectrograph and a controlled, fixed, water vapor path.  For the 8 strong 
H$_2$O lines listed in Table~\ref{tab-strong} that were used for the saturation 
investigation (see \S~\ref{sec-sat}), the $gf$ values were taken from
Coheur et al. (2002) and similar laboratory measurements of Brown, Toth, \& Dulick (2002).
 For the 6943.79~\AA\ line the empirical Coheur et al. (2002) $gf$ value indicated a line 
strength parameter, S, that agreed with the B75 value to within 1\%.  Given the good 
agreement between the experimentally determined S factors, and the  
difficulty in making theoretical $gf$ predictions for weak lines (due to small overlap of
the level wavefunctions), we chose not to use the Kurucz H$_2$O $gf$ values.

\subsection{Results of Optical PWV Measurements}\label{sec-optres}

The range of line equivalent widths, in m\AA , for our list of temperature insensitive 
H$_2$O lines is indicated in Table~\ref{tab-params}.  Table~\ref{tab-pwv} shows the log 
line fluxes and the resultant PWV values computed using Equations~2 and~4 (multiplied by 
10 to get PWV in mm).  The PWV values range from 1 to 7mm for the 15 nights, and there is 
good agreement, typically $\sim$10\%, between the results for each line.  This dispersion 
is consistent with the accuracy of the measured line $gf$ values: good $gf$ values are 
usually accurate to 5 to 10\%.
The weakest line in Table~\ref{tab-params} is at 5954.9~\AA , and is most reliable on nights
with more than $\sim$5mm PWV.  The strongest line in Table~\ref{tab-params}, at 
7287.4~\AA, is useful for measuring PWV on very dry nights; on wet nights it will be the
first line to become saturated.

\begin{figure}
\plotone{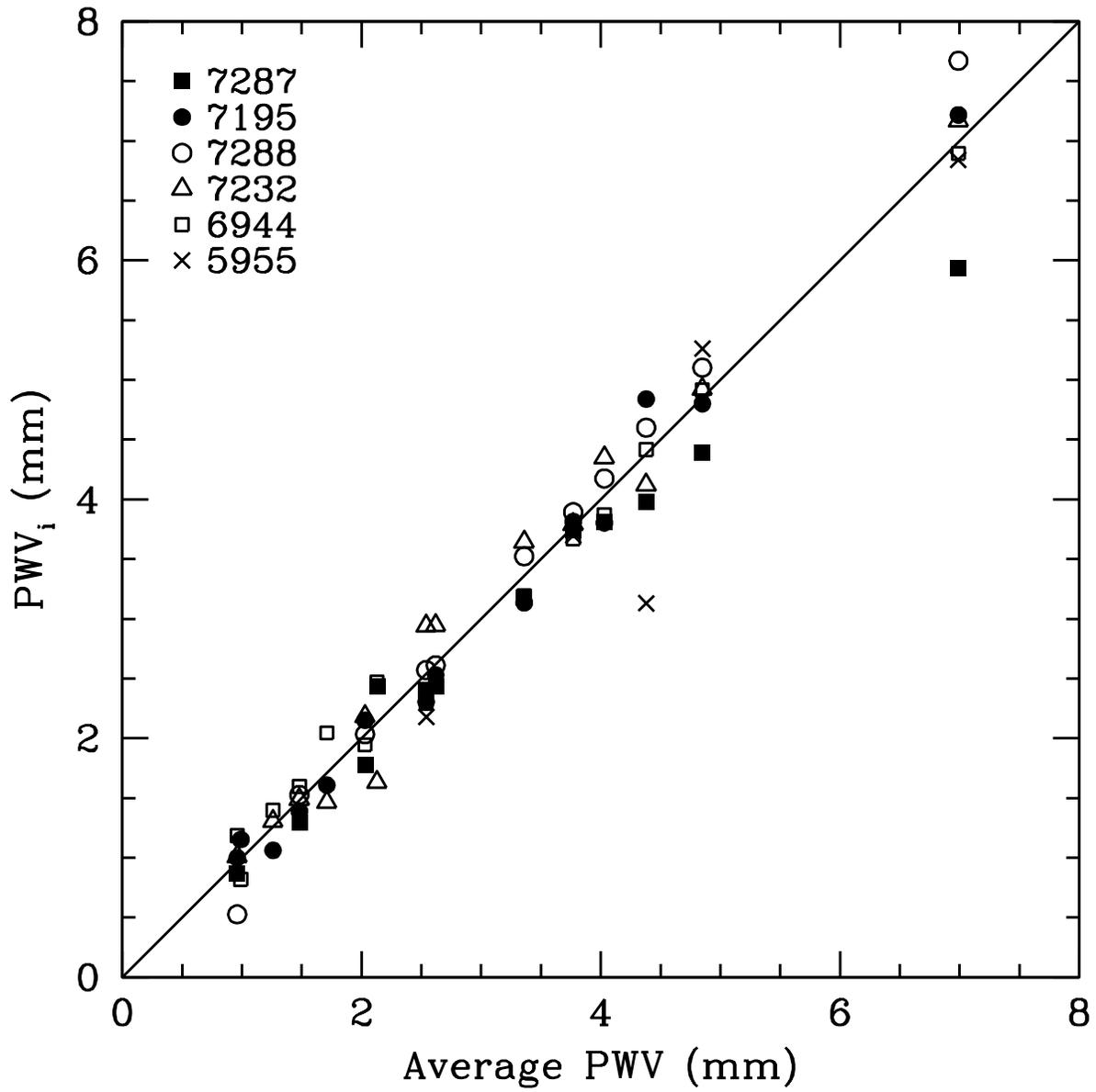}
\caption{A plot showing the individual PWV results from all lines, compared to
the average PWV of each spectrum.  The straight line corresponds to a 1:1 relation.
Symbols for different line wavelengths are indicated top left.
There are no unambiguous systematic trends of individual lines deviating from the mean.}
\label{fig-pwv_comp}
\end{figure}

In Figure~\ref{fig-pwv_comp} we compare the individual PWV results for each line with 
the average
for each spectrum.  The figure shows no clear systematic differences between the PWV
values for the individual lines.  The rms scatter of the individual line PWV measurements 
about the mean for each spectrum is 0.3mm H$_2$O on average, corresponding to $\sim$10\%.  

In Figure~\ref{fig-pwv_comp} the 7287.4~\AA\ point for the wettest night appears below the 
main correlation, slightly more than might be expected from the general scatter of the 
points.  In addition to this for the next two most wettest nights the 7287.4~\AA\ PWV points
are also the lowest of the 6 lines used.  Although these differences are extremely
subtle, and could easily be random, they may indicate the beginning of saturation in
the 7287.4~\AA\ line; indeed, the expected effect of saturation is to reduce the computed
PWV from Equations~2 and~4.  If that is the case then the strength of our 7287.4~\AA\ line on 
the wettest night, at 134 m\AA , represents the useful upper limit for PWV measured using
unsaturated H$_2$O lines under the assumption of no saturation.

The method used to measure PWV here could be applied to sites at higher altitude, 
e.g. Mauna Kea.  The lines will be narrower at higher altitude, due to the lower pressure 
broadening, and thus, saturation will occur at smaller equivalent widths.

\subsection{Saturation Investigation}\label{sec-sat}

As mentioned earlier, the Brault method used to compute PWV in this paper is quite robust,
providing that the lines are not saturated.  To use saturated lines for reliable PWV
measurement would require a detailed model atmosphere treatment where the run of temperature
with PWV, or fractional PWV, is specified through the atmosphere.  Given the small scatter in
our results it seems likely that all of the lines measured in Table~\ref{tab-params}
are unsaturated, although the line at 7287.36~\AA\ may be low on our wettest night, 
indicating the onset of saturation.  The use of saturated lines will result in lower computed
PWV values, with the error becoming more pronounced with increasing line strength.

It would be useful to know the approximate saturation limit in terms of line strength,
so that we do not use overly strong lines in the PWV calculations; this is particularly
important for the wettest nights.  In order to do this we have made a search for strong
H$_2$O lines in the HITRAN database, with lower level excitation energies in the range 
225--300 cm$^{-1}$ (the range that gives PWV results insensitive to atmosphere 
temperature).  We have calculated the parameters for this list of 8 strong H$_2$O lines, 
as shown in Table~\ref{tab-strong}, along with the equivalent widths of each line 
on our wettest night (2 August 2005).  It is interesting that the stronger H$_2$O lines 
tend to lie at redder wavelengths than those in Table~\ref{tab-params}.  

\begin{figure}
\plotone{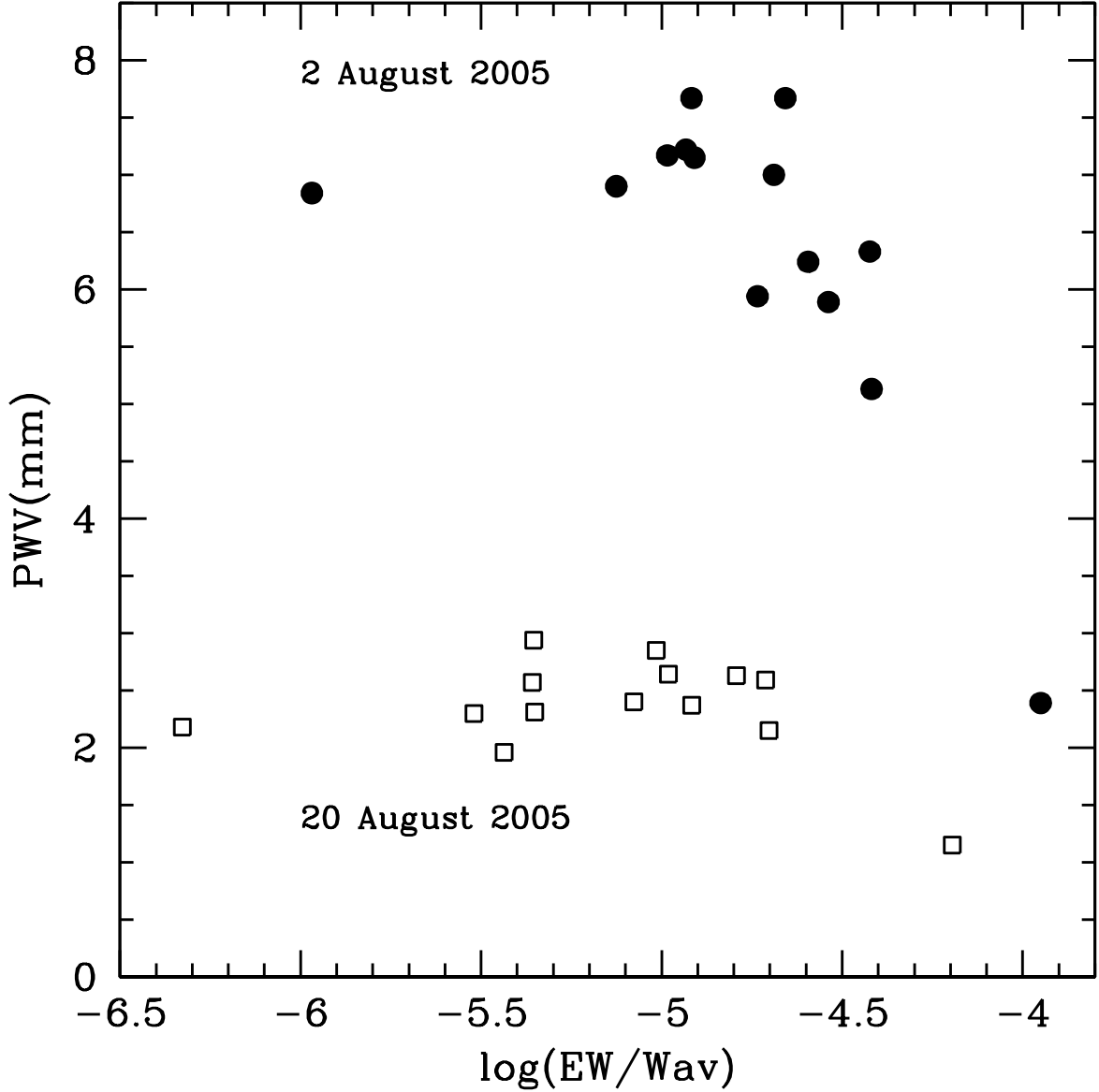}
\caption{A plot showing the results from PWV calculations for strong and weak lines
on the wettest night (2 August 2005, filled circles) and a moderately dry night 
(20 August 2005, open boxes).  For the wettest night there is a clear down-turn in the PWV
estimates above a reduced equivalent width around $-$4.6 to $-$4.7, indicating that lines
above that width suffer from saturation.  On the drier night only one line was saturated;
the excellent agreement between the unsaturated lines shows that the down-turn of the PWV 
points for the wettest night was not due to systematic error in line $gf$ values.}
\label{fig-pwv_sat}
\end{figure}

In Figure~\ref{fig-pwv_sat} we show a plot of reduced equivalent width 
(REW = log$_{10}$ EW/$\lambda$ ) for all 14 H$_2$O lines (from Tables~\ref{tab-params} and
\ref{tab-strong}) versus PWV, for the nights of 2 and 20 August 2005, calculated using 
Equation~2.  It is clear that the stronger lines on 2 August give systematically lower PWV 
values, decreasing roughly linearly with increasing strength; for lines weaker than
REW$\sim$$-$4.6 to $-$4.7 the PWV values are approximately constant.  This saturation limit
corresponds to
$\sim$160~m\AA\ at 7300~\AA ; this is larger than the 134~m\AA\ for the line at 7287.36~\AA\
on the wettest night, which we had previously suspected could be affected by saturation.  
Therefore, future PWV measurements that use the Brault method should disregard lines above
REW$\sim$$-$4.65.  The 20 August points in Figure~\ref{fig-pwv_sat} show that when the 
lines are below the saturation limit they give results in good agreement; this eliminates 
the possibility that the decreasing PWV values for strongest lines on the wettest night are
due to a systematic error in the line $gf$ values.  The unsaturated 20 August points have a 
standard deviation of 11\% in PWV, giving an error on the mean PWV of only 3.3\%.

Because line saturation in red giant stars begins around REW$\sim$$-$5.1 it is clear that
the telluric H$_2$O lines must be more strongly broadened than lines in red giants.
Atmospheres of red giant stars are much hotter and have much larger turbulent velocities 
than the earth's atmosphere; therefore, the  main source of broadening for the telluric 
H$_2$O lines cannot be due to temperature or turbulence.  The only possibility is that the
larger broadening of the telluric lines is due to the much denser earth's atmosphere 
(i.e. pressure broadening), for which one would expect the collision frequency and 
Van der Waals forces to be greater than for lines formed in the low density atmospheres of
stars.  

Given the larger saturation point for telluric lines, than for stellar lines, the pressure
broadening must be overwhelmingly dominant.  Thus, it may be that the saturation point
changes with atmospheric pressure, for example due to the change in the collision frequency
experienced by the H$_2$O molecules.  It is also possible that the saturation point is 
affected by humidity, both because of the presumably larger H$_2$O collision cross-section,
$\sigma$v, and also resulting from different Van der Waals forces for H$_2$O compared to 
other atmospheric gases.

Future PWV measurements can utilize the stronger H$_2$O lines listed in Table~\ref{tab-strong},
provided that the saturation limit is not exceeded.

\subsection{Temperature Diagnostic}

We have attempted to estimate the mean temperature for the wettest night, 2 August 2005, using
H$_2$O lines covering a range of excitation energies.  Iterative calculations were performed
to find the temperature where the line PWVs matched the temperature-independent PWV value.  This
mean temperature is actually weighted by the water vapor mass, so it should reflect the
temperature closer to the ground, where most of the water vapor lies.

We identified and measured nine lines from the HIRTRAN database with energies ranging from 0 to
756 cm$^{-1}$.  The derived temperatures ranged from 268~K to 290~K, with a mean of 279~K and an
rms scatter of 9.5~K.  Some of the scatter may have resulted from blending of weak high
excitation lines.  We note that the mean temperature verifies our use of a temperature of 270~K
for the PWV calculations.

Apart from verifying our initial assumption it is not clear how useful the mean temperature
diagnostic would be; perhaps it may be used to compare with mass-weighted temperatures
calculated from tailored atmosphere models.  A better way to compare with theoretical models 
would be to use profiles of strongly saturated lines to derive the run of temperature with 
H$_2$O opacity; but this would require accurate estimates of the line broadening in the 
theoretical models.

\section{225~GHz Tipping Radiometer Opacity}\label{sec-tip}

Opacity at 225~GHz was studied through the use of the University of Arizona Radio Observatory
(ARO) 225~GHz Tipping Radiometer (Tipper) on loan to Las Campanas Observatory for the 2005 Southern hemisphere winter season.  This radiometer is one of four originally constructed for Millimeter Array site testing purposes by the NRAO and has been in regular use at ARO for sky quality assesment since 1995.  These systems have been used to characterize many other sites including Mauna Kea (one of the units has been in operation since 1989 at the Caltech Submillimeter Observatory), Antarctica, Cerro Chanjnantor in Chile, South Baldy and the VLA site in New Mexico \citep{oh89,sch90,cha04}.

The characteristics of this radiometer have been extensively described elsewhere 
\citep{liu87,mc87,cb94} therefore only a short description is provided here.  The radiometer
is illuminated via a mirror (with a beamwidth of 3.4 degrees) that rotates through 11 elevation
angles between an airmass of one and three thus providing a "tip" (measurement of the sky 
brightness as a function of zenith angle). The beam enters a temperature controlled box in
which a chopper wheel allows the receiver to alternately view in repeatable sequences the sky,
a cold (45~C) load, the sky, and finally a hot (65~C)  load. Use of two different
temperature reference loads allows the gain to be measured.  The opacity and an estimate of its
uncertainty are then derived according to a slab model of the atmosphere from a linear fit to
the natural logarithm of the sky brightness temperature as a function of airmass.  

Opacities were obtained approximately every 5 minutes continuously between 
between July 17, 2005 and August 30, 2005 with some exceptions due to storm 
activity.  Of these data, 2999 measurements were selected from the 29 full 
and partial nights when the conditions were judged to be photometric.  
The requirement of photometric conditions was imposed not only because of 
concerns that non-photometric conditions would lead to incorrect fluxes and 
opacity measurements from the Tipping Radiometer, but also due to the desire 
to characterize the PWV on nights suitable for mid-infrared astronomy.

\subsection{Relationship of 225~GHz opacity to precipitable water vapor}\label{sec-cal}

The relationship between 225~GHz opacity, $\tau$, and the column of precipitable water vapor,
PWV in mm, can be represented as follows:
\begin{equation}
\tau=\tau_{dry-air}+ B \cdot PWV
\end{equation}
where the opacities and the coefficient B are reported in nepers/airmass and nepers/airmass/mm
$H_2O$, respectively.  As used here, a neper is a measurement of attenuation similar to a decibel but based on the natural logarithm, as opposed to base 10.

\citet{pla78} reports the first estimate of the relationship between $\tau$ and PWV.  The 
opacity measurements made at UC Hat Creek Observatory (with an altitude of 1050-m) were 
correlated with PWV as measured by a radiosonde in Medford, OR (210 km to the NW).  Although
no values below 4 mm PWV were measured, the coefficient B was found to be 0.06. In the 
development of the NRAO tipping radiometers, one of which was used in our study, \citet{mc87}
used a similar relationship but included a constant opacity offset of 0.005 nepers/airmass due
to absorption by oxygen.  They also mention that the relationship is weakly dependent on elevation.

\citet{cb95} report on their calibration efforts in Antarctica using another NRAO tipping 
radiometer to measure 225~GHz opacities and Radiosonde upper air soundings to measure PWV.
Their coefficients can be found in Table \ref{tab-cal}.  A re-analysis of this data is 
presented by \citet{cha04} but does not change the finding of a significant non-zero dry air opacity.
The cause of the dry air opacity is not well understood as many atmospheric transmission models
\citep[e.g][]{par01, gro89, lie89} have failed to reproduce it.  A new model that accounts theoretically for the collision-induced absorption by $N_2$-$N_2$ molecular partners may be able to resolve this discprepancy \citep{pai04}.  Further work in understanding and characterizing this parameter is especially important to characterizing the low-precipitable-water-vapor sites at mm wavelengths.

At Mauna Kea this relationship was established in a different way \citep{dav97} and the 
coefficients can also be found in Table \ref{tab-cal}.  The 225~GHz opacities were measured
with the Caltech Submillimeter Observatory (CSO) tipping radiometer, also one of the original
NRAO tipping radiometers.  The PWV values were obtained through the use of an atmospheric
spectral synthesis model called FASCOD2.  As the PWV has not been independently measured, it
is unclear whether this model might also underestimate the dry air opacity.  Given that this
calibration is now canonically used to produce PWV values from 225~GHz opacities for Mauna Kea
as well as to draw comparisons to other sites, it is important to measure the dry air opacity
there.

\begin{figure}
\plotone{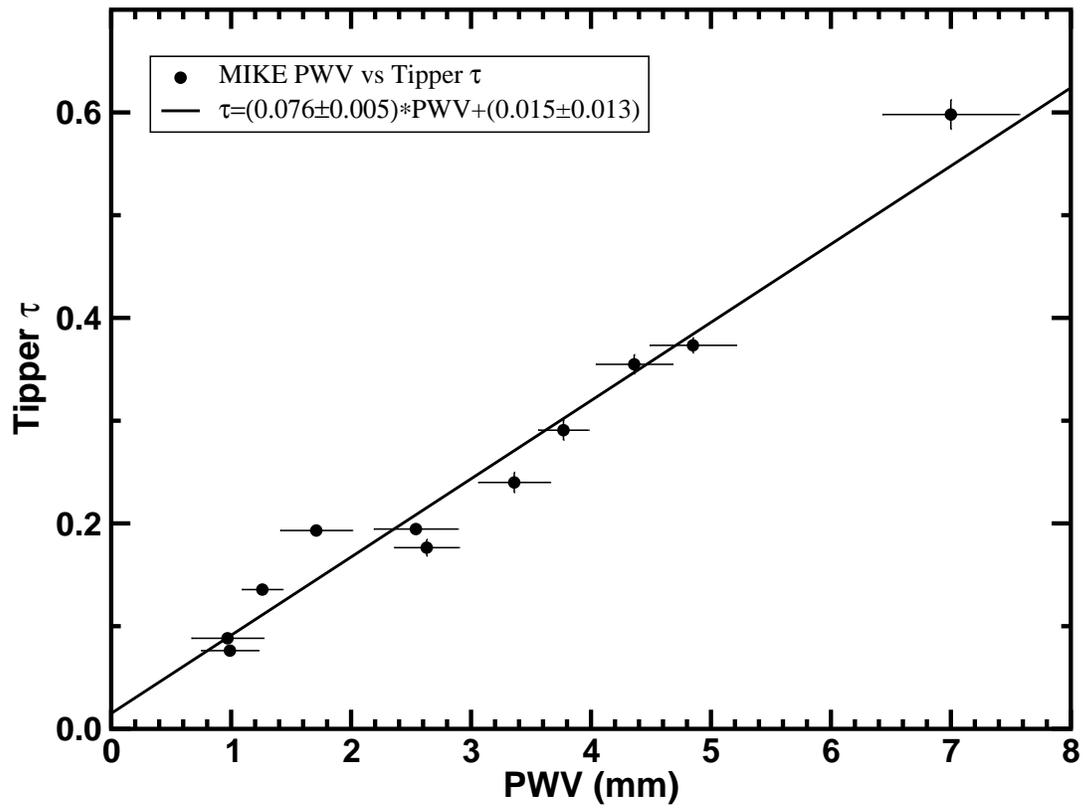}
\caption{The correlation between MIKE echelle PWV Measurements and 225~GHz tipper opacities
at LCO during the period between mid-July and mid-Sept 2005.  Where no error bars are 
visible the errors are within the size of the symbol.  The line represents a linear least 
squares fit with characteristics denoted in the legend.}
\label{fig-pwv_tau}
\end{figure}

At Las Campanas Observatory, the precipitable water vapor values determined with the MIKE 
echelle spectrograph in Section \ref{sec-optres} and shown in Table \ref{tab-pwv} were 
correlated with the 225~GHz opacity measurements as shown in Figure \ref{fig-pwv_tau}.  The
opacities at the nearest times to those in Table \ref{tab-pwv} were selected.  There was 
never more than 5 minutes difference and a total of 11 points with times in common were found.
A linear relationship was determined using a linear least squares regression with errors in
both the independent and dependent variables.  Our coefficients are also listed in 
Table \ref{tab-cal}.  More points would be needed to better constrain the dry air opacity.

A meaningful comparison of the available calibration coefficients at different sites is made
difficult by the lack of uncertainties reported for the calibrations cited in 
Table \ref{tab-cal}.  \citet{cb95} reported very small uncertainties for their calibration 
efforts at the South Pole.  Their results, however, were superceded by a re-analysis of the
tipping radiometer measurements in which the opacities were adjusted systematically upwards
by 15\% \citep{cha04}.  No mention of the relative uncertainties was made in \citet{cha04}
and since adjustments were systematic in nature we will quote the original uncertainty values from \citet{cb95}
in Table \ref{tab-cal}.  Regardless of the lack of uncertainties in the reported 
calibrations, it appears that the dry air opacity decreases with altitude and as expected 
due to the pressure broadening seen in the wing of the 183~GHz line, the coefficient B, 
also decreases with altitude.  In other words, at lower altitudes (higher pressure) less 
PWV is required to produce the same opacity at 225 GHz as seen at higher altitudes.

\subsection{Calibrated precipitable water vapor at Las Campanas Observatory}\label{sec-pwv}

Given the calibration determined in the previous section, millimeters of precipitable water
vapor and its uncertainty estimate are calculated from the tipper opacities and errors in 
the following manner:

\begin{eqnarray}
PWV=\frac{\tau - \tau_{dry-air}}{B}\\
\sigma_{pwv}=\sqrt{\sigma_{\tau_{dry-air}}^2+(\tau \cdot \sigma_B)^2+B \cdot \sigma_{\tau}^2}\label{eq:pwvcal}
\end{eqnarray}

\begin{figure}
\plotone{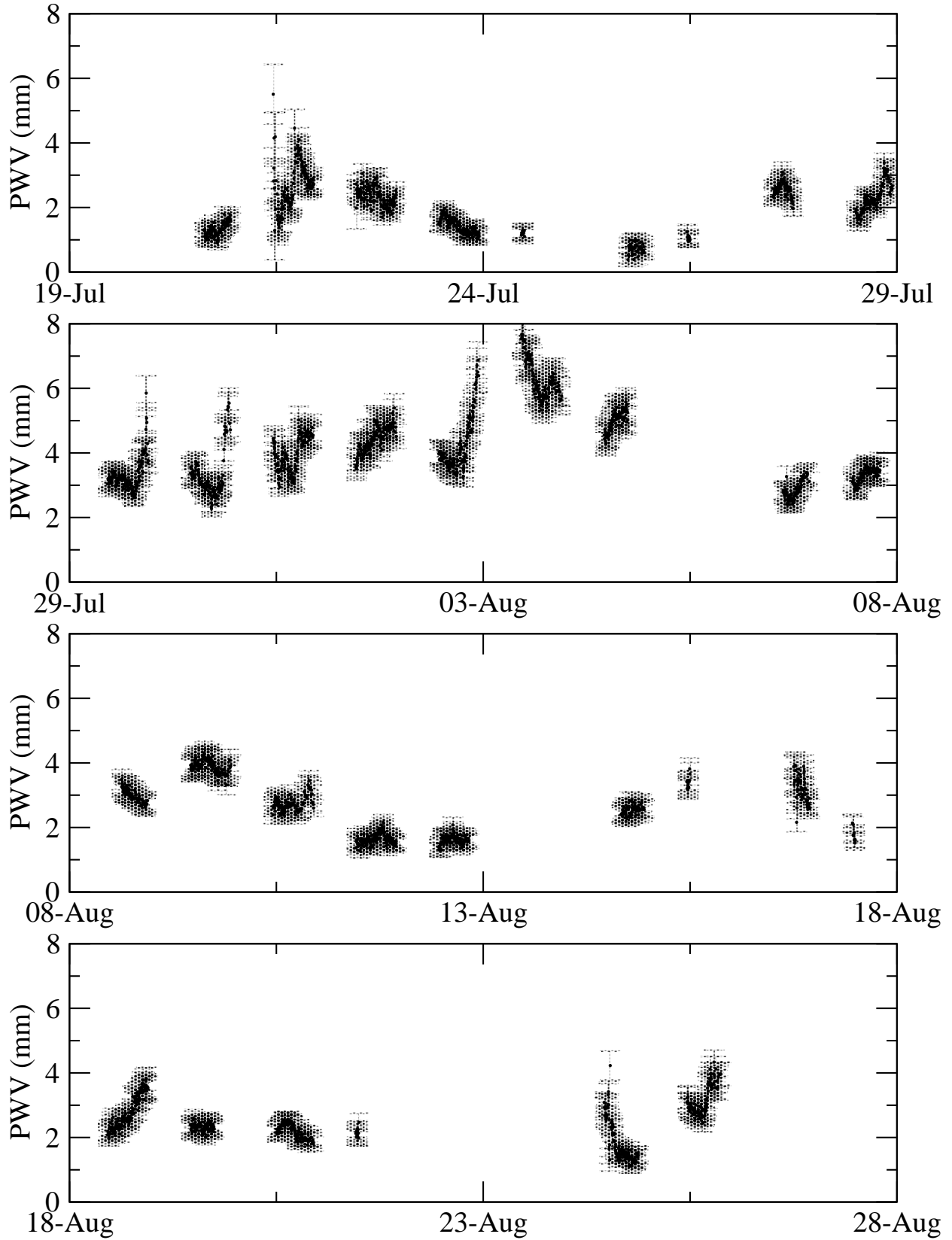}
\caption{Clear night-time PWV as a function of time during our campaign.  Uncertainties are
indicated in gray.}
\label{fig-tippwv}
\end{figure}

\begin{figure}
\plotone{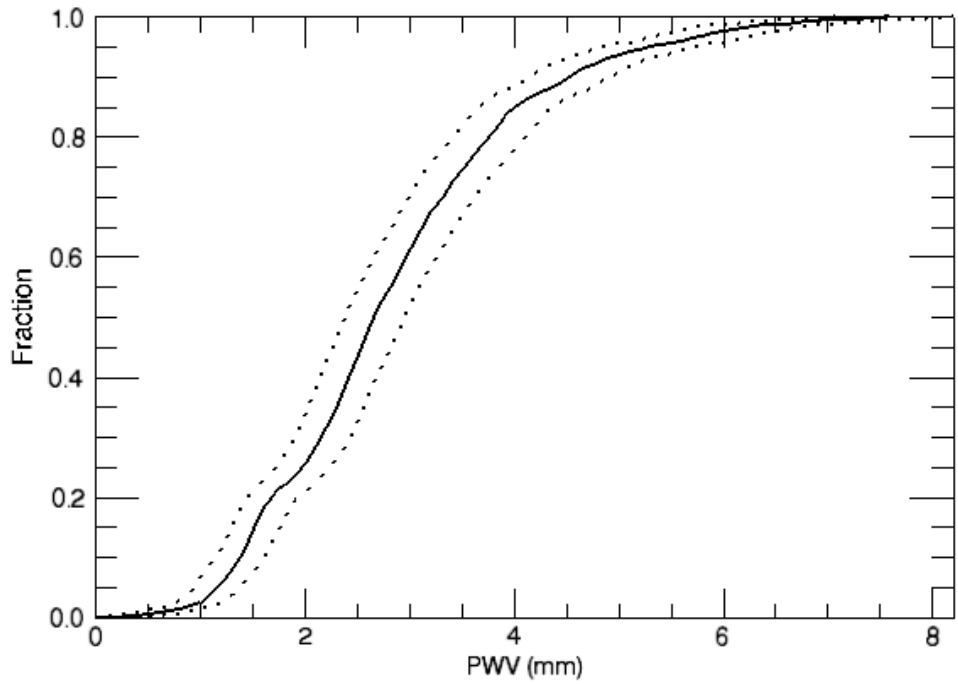}
\caption{Solid line represents the fraction of PWV measurements below a given value.  Dotted
lines represent the uncertainties due to both the measurements and the calibration.  The 10,
25, 50, 75, and 90 percentiles are 1.5, 2.1, 2.8, 3.6, and 4.6 mm, respectively.}
\label{fig-tippwvhist}
\end{figure}

Figure \ref{fig-tippwv} shows the resulting precipitable water vapor values at night under
photometric conditions as a function of time during our campaign.  The fraction of the 
measurements for which PWV was found to be below a given value is displayed in 
Figure \ref{fig-tippwvhist}.  Uncertainties in our PWV estimates are on the order of 10\%.

Over the 1.5 month period covered by our measurements, we find a median 
PWV of 2.8$\pm$0.3 mm.  Assuming that this time period is representative of 
the winter PWV characteristics at Las Campanas Observatory, then we can 
expect good 
conditions for mid-infrared observing ($<$1.5 mm PWV) 10\% of the time in 
winter \citep[c.f.][]{gio01}.  These results
agree well with measurements made at the nearby La Silla Observatory during 
the VLT site survey \citep{swi90}.  La Silla is located 24~km south of Las
Campanas, and is nearly exactly the same altitude.  The VLT site survey 
measurements were obtained using a mid-infrared sky radiance monitor 
\citep{mg82}.  The absolute scale of these data was calibrated using the
B75 method on coud\'{e} spectra of the 6943.79~\AA\ line obtained
simultaneously with mid-infrared sky radiance measurements on several 
nights.

\section{Conclusions}

We report on the calibration of the relationship between the column of precipitable water 
vapor, PWV, and opacity at 225~GHz at Las Campanas Observatory as measured by an NRAO 
Tipping Radiometer on loan from the University of Arizona Radio Observatory and the high 
resolution echelle spectrograph, MIKE, on the Magellan Clay Telescope.  We have expanded 
the absolutely calibrated method for measuring PWV using temperature insensitive lines in 
high resolutions stellar spectra presented by \citet{b75} with improved partition functions
and 13 additional lines.  We found that optical H$_2$O lines with reduced equivalent
widths weaker than $\sim$$-$4.7 to $-$4.6 are unsaturated and suitable for measuring PWV
using the Brault method.  This saturation level indicates that the lines are strongly
pressure broadened.  A calculation of the mass-weighted mean temperature validates the
use of a single slab model atmosphere with T=270~K to compute PWV values.  The linear
relationship we found between the MIKE PWV and the 225~GHz opacities is consistent with 
opacities formed from a pressure broadened wing of the strong water vapor line at 183~GHz 
and including a dry-air opacity component.  The effect of altitude (due to atmospheric 
pressure changes) is demonstrated by comparing our calibration at Las Campanas Observatory 
with those few available in the literature.  We note the paucity of such absolute PWV 
calibration efforts and encourage further work especially at sites like Mauna Kea and 
Paranal where the capability for high resolution echelle spectra already exists.

Based on the relationship determined between the measured PWV and 225~GHz 
opacities, we find a median PWV of 2.8$\pm$0.3 mm and a range between 0.5~mm 
and 7.5~mm for our two month Southern hemisphere winter time campaign.  This 
is consistent with that measured at the nearby La Silla Observatory during 
the VLT site survey \citet{swi90}.  Finally, if our campaign is representative 
of the Southern hemisphere winter months, we can expect good conditions for 
infrared observing ($\lesssim$1.5 mm) at the tenth percentile level.

\acknowledgments

We are grateful to Lucy Ziurys and the Arizona Radio Observatories for the loan of  the Tipping Radiometer.   We also appreciate Mike Meyer's help in arranging for the loan.  A.M. gratefully acknowledges support from NSF grants AST-96-18623 and AST-00-98612 and I.T. from NSF grant AST-0507325.   J.T-O. thanks John Bally for a particularly fruitful conversation regarding PWV calibration efforts and John Grula for his superb efforts in tracking down obscure references.  Finally, we are indebted to Magellan/MIKE observers, Andrea Dupree, Varsha Kulkarni, James Lauroesch, and George Preston, for obtaining spectra.


{\it Facilities:} \facility{Magellan (MIKE)}, \facility{ARO (Tipper)}.

\clearpage

\begin{deluxetable}{rrrcccl}
\tablenum{1}
\tablewidth{0pt}
\tablecaption{Log of Observations}
\tablehead{\colhead{Obs. \#} & \colhead{HR \#} & \colhead{UT Date} & \colhead{UT Start} & \colhead{Airmass} & \colhead{Exp. (sec)} &
\colhead{Observer\tablenotemark{*} } }
\startdata
1   & 5987  & 21/Jul/2005 &  22:49  &  1.116 & 24 & ALM   \cr
2   & 5987  & 22/Jul/2005 &  22:46  &  1.115 & 32 & ALM   \cr
3   &  806  & 24/Jul/2005 &  11:01  &  1.292 & 30 & ALM   \cr
4   & 5987  & 25/Jul/2005 &  01:37  &  1.033 & \hskip6pt 5 & AD    \cr
5   & 845   & 30/Jul/2005 &  10:42  &  1.002 & 15 & IT    \cr
6   & 5517  & 30/Jul/2005 &  22:44  &  1.002 & 15 & IT    \cr
7   & 5517  &  1/Aug/2005 &  22:48  &  1.001 & 15 & IT    \cr
8   & 5517  &  2/Aug/2005 &  23:08  &  1.005 & 15 & IT    \cr
9   & 6930  & 20/Aug/2005 &  01:52  &  1.043 & 15 & GWP   \cr
10  & 5987  & 20/Aug/2005 &  23:10  &  1.011 & 32 & GWP   \cr
11  & 5987  & 23/Aug/2005 &  23:08  &  1.014 & 62 & GWP   \cr
12  & 5987  & 24/Aug/2005 &  22:52  &  1.011 & 30 & GWP   \cr
13  & 6930  & 15/Sep/2005 &  01:00  &  1.098 & 10 & KL    \cr
14  & 8998  & 21/Nov/2005 &  23:45  &  1.030 & 49 & AMcW  \cr
15  & 3034  & 22/Nov/2005 &  08:21  &  1.002 & \hskip6pt 3 & NM    \cr

\enddata
\tablenotetext{*}{Observer key: AD (Andrea Dupree), IT (Ian Thompson), GWP (George Preston),
KL (Kulkarni \& Lauroesch), AMcW (Andrew McWilliam), NM (Nidia Morel), 
ALM (Adams \& Lopez Moralez)}
\label{tab-log}
\end{deluxetable}

\clearpage

\begin{deluxetable}{cccllcc}
\tablenum{2}
\tablewidth{0pt}
\tablecaption{Parameters of Temperature Insensitive H$_2$O Lines\tablenotemark{*}}
\tablehead{\colhead{$\sigma_{\rm vac}$}  & \colhead{$\lambda_{\rm air}$}
        &  \colhead{EP}                 & \colhead{gf} 
       &  \colhead{S$_{270}$}  & \colhead{EW$_{dry}$}   & \colhead{EW$_{wet}$} \cr

\colhead{(cm$^{-1}$)}  & \colhead{(\AA )}
        &  \colhead{(cm$^{-1}$)}                 & 
       &  \colhead{(cm$^{-2}$)}  & \colhead{(m\AA )}  & \colhead{(m\AA )} 
}
\startdata
13717.1744  &  7288.10  &  275.50  & 5.323$\times$10$^{-9}$   & 0.2387 &  6 &  88 \cr
13718.5755  &  7287.36  &  300.36  & 1.251$\times$10$^{-8}$   & 0.4912 & \llap{2}1 & \llap{1}34 \cr
13823.1810  &  7232.21  &  300.36  & 5.602$\times$10$^{-9}$   & 0.2200 &  9 &  75 \cr
13894.6353  &  7195.02  &  285.42  & 6.159$\times$10$^{-9}$   & 0.2619 & \llap{1}2 &  84 \cr
14397.3641  &  6943.79  &  224.84  & 3.059$\times$10$^{-9}$   & 0.1795\rlap{\tablenotemark{**}}  & 5\rlap{:} &  52 \cr
16788.1104  &  5954.94  &  300.36  & 8.942$\times$10$^{-10}$  & 0.03511 & ... &  \hskip6pt 6 \cr
\enddata
\tablenotetext{*}{S and gf values calculated from Coheur et al. (2002) A values.}
\tablenotetext{**}{Compares well with Brault et al. (1975) S value of 0.180.}
\label{tab-params}
\end{deluxetable}

\clearpage

\begin{deluxetable}{r|cc|cc|cc|cc|cc|cc|cc}
\tabletypesize{\tiny}
\rotate
\tablenum{3}
\tablecolumns{15}
\tablewidth{0pt}
\tablecaption{Line Fluxes\tablenotemark{*} and PWV Values}
\tablehead{
\colhead{Obs.}  &
\multicolumn{2}{c}{7288.10\AA}   &        \multicolumn{2}{c}{7287.36\AA}       &  
          \multicolumn{2}{c}{7232.21\AA}     &  \multicolumn{2}{c}{7195.02\AA}      &     \multicolumn{2}{c}{6943.78\AA}      &
          \multicolumn{2}{c}{5954.94\AA\tablenotemark{\ddag}}     &  \colhead{$\overline{\rm PWV}$\tablenotemark{\dag} (mm)} & \colhead{$\sigma$}  \cr
     &     \colhead{$-$ln F} &  \colhead{PWV} &  \colhead{$-$ln F} &  \colhead{PWV} &  \colhead{$-$ln F} &  \colhead{PWV} &
           \colhead{$-$ln F} &  \colhead{PWV} &  \colhead{$-$ln F} &  \colhead{PWV} &  \colhead{$-$ln F} &  \colhead{PWV} & & }
\startdata
 1   &    ...   &  ...   &    ...   &  ...  &   0.036\rlap{:} & 1.47  &  0.047\rlap{:} & 1.61  &    0.041   & 2.05  &      ...    &   ... &     1.71  &  0.30  \cr
 2   &    ...   &  ...   &    ...   &  ...  &   0.032\rlap{:} & 1.31  &  0.031\rlap{:} & 1.06  &    0.028   & 1.40  &      ...    &   ... &     1.26  &  0.17  \cr
 3   &    ...   &  ...   &    ...   &  ...  &    ...   &  ...  &  0.039\rlap{:} & 1.15  &    0.019\rlap{:}  & 0.82  &      ...    &   ... &     0.99  &  0.24  \cr
 4   &   0.013\rlap{:} & 0.53   &   0.044  & 0.87  &   0.023\rlap{:} & 1.01  &  0.027\rlap{7} & 1.00  &    0.027   & 1.19  &      ...    &   ... &     0.97  &  0.30  \cr
     &          &        &          &       &          &       &         &       &    0.017\rlap{:}  &       &             &       &           &        \cr
 5   &   0.122  & 5.10   &   0.216  & 4.39  &   0.116  & 4.92  &  0.126  & 4.80  &    0.095   & 4.92  &     0.022\rlap{:}  &  5.26 &     4.85  &  0.36  \cr
     &          &        &          &       &   0.101  &       &         &       &    0.082   &       &     0.015   &       &           &        \cr
 6   &   0.110  & 4.60   &   0.196  & 3.98  &   0.098  & 4.12  &  0.127  & 4.84  &    0.079   & 4.42  &     0.011\rlap{:}  &  3.13 &     4.36  &  0.32  \cr
     &          &        &          &       &   0.087  &       &         &       &    0.080   &       &     0.011   &       &           &        \cr
 7   &   0.093  & 3.89   &   0.184  & 3.74  &   0.085  & 3.79  &  0.100  & 3.81  &    0.060   & 3.68  &     0.013   &  3.70 &     3.77  &  0.21  \cr
     &          &        &          &       &   0.082  &       &         &       &    0.072   &       &             &       &           &        \cr
 8   &   0.184  & 7.67   &   0.293  & 5.94  &   0.166  & 7.17  &  0.190  & 7.22  &    0.127   & 6.90  &     0.295\rlap{:}  &  6.84 &     7.00  &  0.57  \cr
     &          &        &          &       &   0.151  &       &         &       &    0.122   &       &     0.188   &       &           &        \cr
 9   &   0.064  & 2.57   &   0.123  & 2.40  &   0.063  & 2.94  &  0.063  & 2.31  &    0.048   & 2.30  &     0.008\rlap{::} &  2.18 &     2.54  &  0.35  \cr
     &          &        &          &       &   0.072  &       &         &       &    0.038   &       &             &       &           &        \cr
10   &   0.063  & 2.61   &   0.121  & 2.44  &   0.065  & 2.95  &  0.067  & 2.53  &    0.049   & 2.45  &      ...    &   ... &     2.63  &  0.27  \cr
     &          &        &          &       &   0.066  &       &         &       &    0.040   &       &             &       &           &        \cr
11   &   0.101  & 4.17   &   0.190  & 3.81  &   0.095  & 4.35  &  0.101  & 3.80  &    0.073   & 3.87  &      ...    &   ... &     4.03  &  0.27  \cr
     &          &        &          &       &   0.099  &       &         &       &    0.068   &       &             &       &           &        \cr
12   &   0.085  & 3.52   &   0.158  & 3.18  &   0.075  & 3.64  &  0.083  & 3.14  &    0.061   & 3.20  &      ...    &   ... &     3.36  &  0.30  \cr
     &          &        &          &       &   0.087  &       &         &       &    0.055   &       &             &       &           &        \cr
13   &   0.040  & 1.53   &   0.076  & 1.30  &   0.036  & 1.49  &  0.040  & 1.39  &    0.033   & 1.60  &      ...    &   ... &     1.49  &  0.12  \cr
     &          &        &          &       &   0.036  &       &         &       &    0.030   &       &             &       &           &        \cr
14   &    ...   &  ...   &   0.122\rlap{:} & 2.44  &   0.034\rlap{:} & 1.63  &   ...   &  ...  &    0.037   & 2.40  &      ...    &   ... &     2.10  &  0.54  \cr
     &          &        &          &       &   0.038  &       &         &       &    0.052\rlap{:}  &       &             &       &           &        \cr
15   &   0.050  & 2.03   &   0.094  & 1.78  &   0.052  & 2.19  &  0.058  & 2.15  &    0.036   & 2.00  &      ...    &   ... &     2.05  &  0.17  \cr
     &          &        &          &       &   0.047  &       &         &       &    0.036   &       &             &       &           &        \cr
\enddata
\label{tab-pwv}
\tablenotetext{*}{\ Flux measurements for lines appearing in a second order are indicated below the first entry}
\tablenotetext{\dag}{\ Lines appearing in two orders were counted twice for the calculation of the average PWV}
\tablenotetext{\ddag}{\ The 5954.94\AA\ line was not used for the average PWV}
\end{deluxetable}

\clearpage

\begin{deluxetable}{cccllr}
\tablenum{4}
\tablewidth{0pt}
\tablecaption{Parameters for Strong H$_2$O Lines}
\tablehead{\colhead{$\sigma_{\rm vac}$}  & \colhead{$\lambda_{\rm air}$}
        &  \colhead{EP}                 & \colhead{gf} 
       &  \colhead{S$_{270}$}  & \colhead{EW$_{wet}$} \cr

\colhead{(cm$^{-1}$)}  & \colhead{(\AA )}
        &  \colhead{(cm$^{-1}$)}                 & 
       &  \colhead{(cm$^{-2}$)}  & \colhead{(m\AA )} 
}
\startdata
10704.4205  &  9339.33  & 300.36   & 2.730$\times$10$^{-7}$   &  10.7240\tablenotemark{2}  &  1047. \cr
11002.2178  &  9086.55  & 285.22   & 1.245$\times$10$^{-8}$   &   0.5303\tablenotemark{2}  &   186. \cr
11124.6353  &  8986.56  & 300.36   & 3.017$\times$10$^{-8}$   &   1.1852\tablenotemark{2}  &   343. \cr
12254.5139  &  8157.99  & 300.36   & 1.676$\times$10$^{-8}$   &   0.6584\tablenotemark{1}  &   208. \cr
11096.9124  &  9009.01  & 285.22   & 4.443$\times$10$^{-9}$   &   0.1892\tablenotemark{2}  &   111. \cr
12238.3078  &  8168.79  & 222.05   & 7.226$\times$10$^{-9}$   &   0.4309\tablenotemark{1}  &   180. \cr
12062.4127  &  8287.91  & 224.84   & 1.770$\times$10$^{-8}$   &   1.0397\tablenotemark{1}  &   313. \cr
12037.5135  &  8305.05  & 325.35   & 2.243$\times$10$^{-8}$   &   0.7714\tablenotemark{1}  &   241. \cr
\enddata
\tablenotetext{1}{S and gf values calculated from Coheur et al. (2002) A values.}
\tablenotetext{2}{S and gf values calculated from Brown et al. (2002) A values.}
\label{tab-strong}
\end{deluxetable}

\clearpage

\begin{deluxetable}{lrrrl}
\tablenum{5}
\tablewidth{0pt}
\tablecaption{Comparison of calibration at LCO to that at other sites}
\tablehead{\colhead{Site} & \colhead{Altitude (km)} & \colhead{$\tau_{dry-air}$} & \colhead{B} & \colhead{Reference}}
\startdata
LCO Alcaino  & 2410  & 0.015$\pm$0.013 &  0.076$\pm$0.005  &  this work \cr
S. Pole  & 2835  & 0.026$\pm$0.001\tablenotemark{*} &  0.083$\pm$0.002\tablenotemark{*}  &  \citet{cha04} \cr
Mauna Kea & 4100 & 0.016 & 0.05 & \citet{dav97} \cr
Llanno de Chanjantor & 5104 & 0.0068 & 0.0407 & \citet{del99} \cr
\enddata
\tablenotetext{*}{uncertainties from \citet{cb95}}
\label{tab-cal}
\end{deluxetable}

\clearpage

\clearpage

\clearpage

\clearpage

\clearpage

\clearpage

\clearpage

\clearpage

\clearpage

\end{document}